\documentclass[twocolumn]{aastex63}

\shorttitle{Rapid bursts of accretion in V1025 Cen}
\shortauthors{Littlefield et al.}

\submitjournal{ApJL}
\accepted{December 7, 2021}

\newcommand{\tess}{\textit{TESS}}

\begin{document}

\title{Rapid bursts of magnetically gated accretion in the intermediate polar V1025 Cen }
\author{Colin Littlefield}
\affiliation{Department of Physics, University of Notre Dame, Notre Dame, IN 46556, USA}
\affiliation{Department of Astronomy, University of Washington, Seattle, WA 98195, USA}
\affiliation{Bay Area Environmental Research Institute, Moffett Field, CA 94035, USA}

\author{Jean-Pierre Lasota}
\affiliation{Institut d'Astrophysique de Paris, CNRS et Sorbonne Universit\'e, UMR 7095, 98bis Bd Arago, 75014, Paris, France}
\affiliation{Nicolaus Copernicus Astronomical Center, Polish Academy of Sciences, Bartycka 18, 00-716, Warsaw, Poland}

\author{Jean-Marie Hameury}
\affiliation{Observatoire Astronomique de Strasbourg, Universit\'e de Strasbourg, CNRS UMR 7550, 67000, Strasbourg, France}

\author{Simone Scaringi}
\affiliation{Centre for Extragalactic Astronomy, Department of Physics, University of Durham, South Road, Durham DH1 3LE, UK}

\author{Peter Garnavich}
\affiliation{Department of Physics, University of Notre Dame, Notre Dame, IN 46556, USA}

\author{Paula Szkody}
\affiliation{Department of Astronomy, University of Washington, Seattle, WA 98195, USA}

\author{Mark Kennedy}
\affiliation{Department of Physics, University College Cork, Cork, Ireland}

\author{McKenna Leichty}
\affiliation{Department of Physics, University of Notre Dame, Notre Dame, IN 46556, USA}

\correspondingauthor{Colin Littlefield}
\email{clittlef@alumni.nd.edu}

\begin{abstract}

   Magnetically gated accretion has emerged as a proposed mechanism for producing extremely short, repetitive bursts of accretion onto magnetized white dwarfs in intermediate polars (IPs), but this phenomenon has not been detected previously in a confirmed IP. We report the 27~d \textit{TESS} light curve of V1025~Cen, an IP that shows a remarkable series of twelve bursts of accretion, each lasting for less than six hours. The extreme brevity of the bursts and their short recurrence times ($\sim1-3$~d) are incompatible with the dwarf-nova instability, but they are natural consequences of the magnetic gating mechanism developed by Spruit \& Taam to explain the Type~II bursts of the accreting neutron star known as the Rapid Burster. In this model, the accretion flow piles up at the magnetospheric boundary and presses inward until it couples with the star's magnetic field, producing an abrupt burst of accretion. After each burst, the reservoir of matter at the edge of the magnetosphere is replenished, leading to cyclical bursts of accretion. A pair of recent studies applied this instability to the suspected IPs MV~Lyr and TW~Pic, but the magnetic nature of these two systems has not been independently confirmed. In contrast, previous studies have unambiguously established the white dwarf in V1025~Cen to be significantly magnetized. The detection of magnetically gated bursts in a confirmed IP therefore validates the extension of the Spruit \& Taam instability to magnetized white dwarfs.
   \\
   \\
    
\end{abstract}


\section{Introduction}

\begin{figure*}
    \centering
    \includegraphics[width=\textwidth]{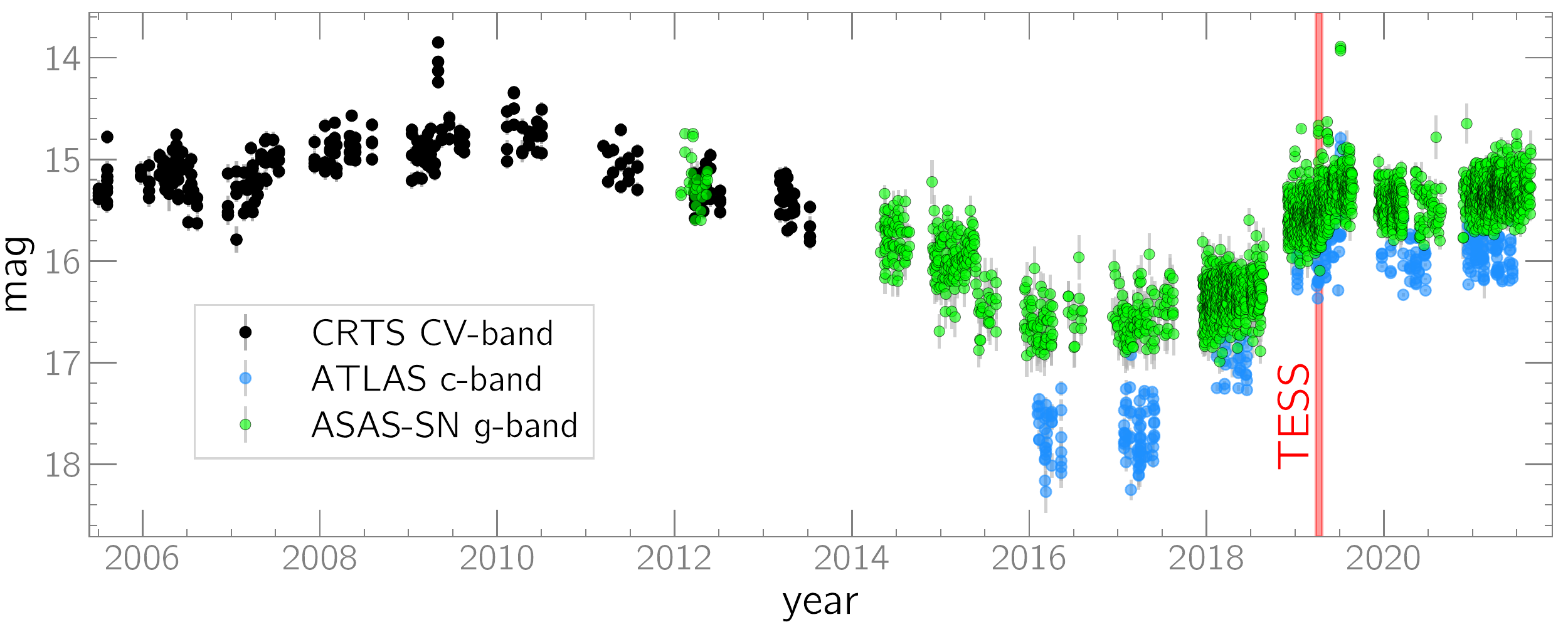}
    \caption{ The combined CRTS, ASAS-SN, and ATLAS light curve of V1025 Cen. The timespan of the \tess\ observations is indicated in red. V1025~Cen experienced a low-accretion state between 2015 and 2018, and we believe that the unblended ATLAS data offer a better measurement of its depth compared to the blended ASAS-SN data. A single outburst, with a minimum amplitude of 0.6~mag, was detected by ASAS-SN during the \tess\ observations. There were several additional bursts in the months before and after the \tess\ observation, and the CRTS data show a single burst of unknown duration in 2009.
    }
    \label{fig:asassn}
\end{figure*}

\begin{figure*}
    \centering
    \includegraphics[width=\textwidth]{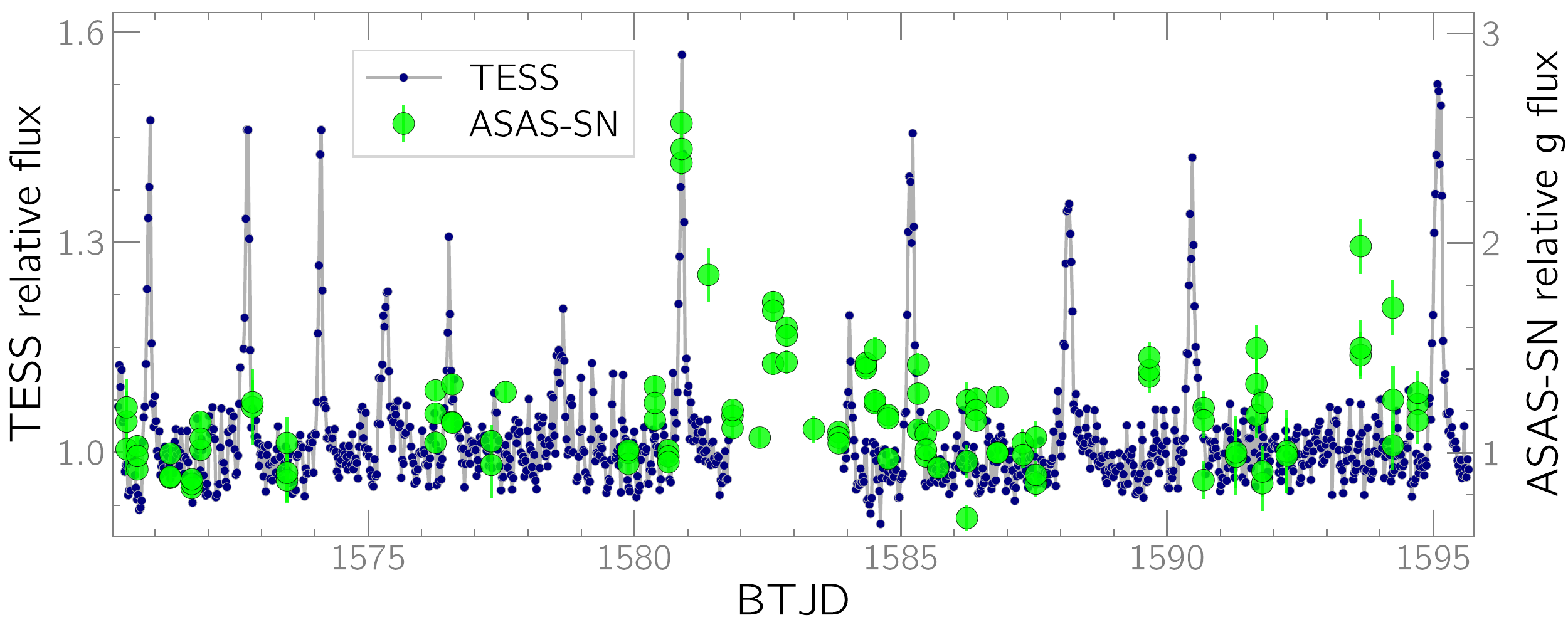}
    \caption{\tess\ light curve of V1025 Cen, with simultaneous ASAS-SN observations overlaid on the secondary $y$-axis. Each \tess\ datum is a 30~min integration, while the ASAS-SN observations are 90~s exposures.
    }
    \label{fig:TESS_LC}
\end{figure*}

    Accreting, non-magnetic white dwarfs (WDs) in cataclysmic variable stars (CVs) occasionally undergo dwarf-nova outbursts, which are brief episodes of greatly enhanced accretion. The mechanism that produces these outbursts, the dwarf-nova instability (DNI), is the runaway ionization of the accretion disk when it exceeds a critical effective temperature \citep[for reviews, see][]{osaki,lasota,hameury}. As it ionizes, the disk's temperature and viscosity skyrocket, resulting in a temporary increase in the accretion rate onto the WD. A typical outburst will result in a multi-magnitude brightening in optical photometry for several days. 
    
    Despite the many successes of the DNI model, it has struggled to explain the properties of outbursts observed in intermediate polars (IPs), the subset of CVs that contains WDs whose magnetic-field strengths are high enough to disrupt the accretion flow without synchronizing the WD's spin to the binary orbital period \citep[for a review, see][]{patterson}. Although some IPs, such as CC~Scl, exhibit bona fide dwarf-nova outbursts, outbursts in other IPs can last for less than one day, with long, irregular recurrence intervals \citep[Table~1 in][]{hl17}; their amplitudes also tend to be lower, typically less than about $\sim1-2$~mag. \citet{hl17} concluded that the DNI cannot explain the very short ($\sim$hours-long) outbursts observed in some IPs, implying that a different mechanism is responsible for very short IP outbursts. 
    
    Recently, \citet{scaringi, scaringi21} identified a repetitive series of rapid, low-amplitude bursts of accretion in the nominally non-magnetic systems MV~Lyr and TW~Pic, respectively. Both studies attributed these bursts to magnetically gated accretion via the \citet{st93} instability, which occurs in systems whose inner disk radius, defined as the magnetospheric radius, is close to the corotation radius. In such a situation, the rapidly rotating magnetosphere creates a centrifugal barrier that suppresses accretion and causes the accretion flow to pile up just outside the magnetospheric boundary. Eventually, this material compresses the magnetosphere until it is able to couple onto magnetic field lines and accrete. Once the reservoir of matter outside the magnetosphere is depleted, the cycle repeats itself, giving rise to episodic spurts of accretion. \citet{scaringi, scaringi21} determined that this process would occur at a critical mass-transfer rate determined by the WD's magnetic-field strength and its rotational frequency. However, neither of these two systems was previously known to possess a magnetized WD, and the bursts themselves provided the main evidence for the WD's magnetism.

    \subsection{V1025 Cen}
    
    V1025~Cen was identified as an IP by \citet{buckley98} with optical photometry and spectroscopy and proposed an unusually long spin period ($P_{spin} = 36$~min) relative to the binary orbital period ($P_{orb} = 85$~min). \citet{hellier98} confirmed this classification and the periods with X-ray observations. More recently, \citet{hellier02} presented phase-resolved optical spectroscopy in an effort to ascertain whether V1025~Cen accretes from a truncated accretion disk or is instead diskless, but the observations did not clearly favor either scenario. 
    

    \citet{BJ21} estimate a geometric distance of $196\pm3$~pc based on their analysis of Gaia EDR3 \citep{gaia, edr3}.

\begin{figure*}
    \centering
    \includegraphics[width=\textwidth]{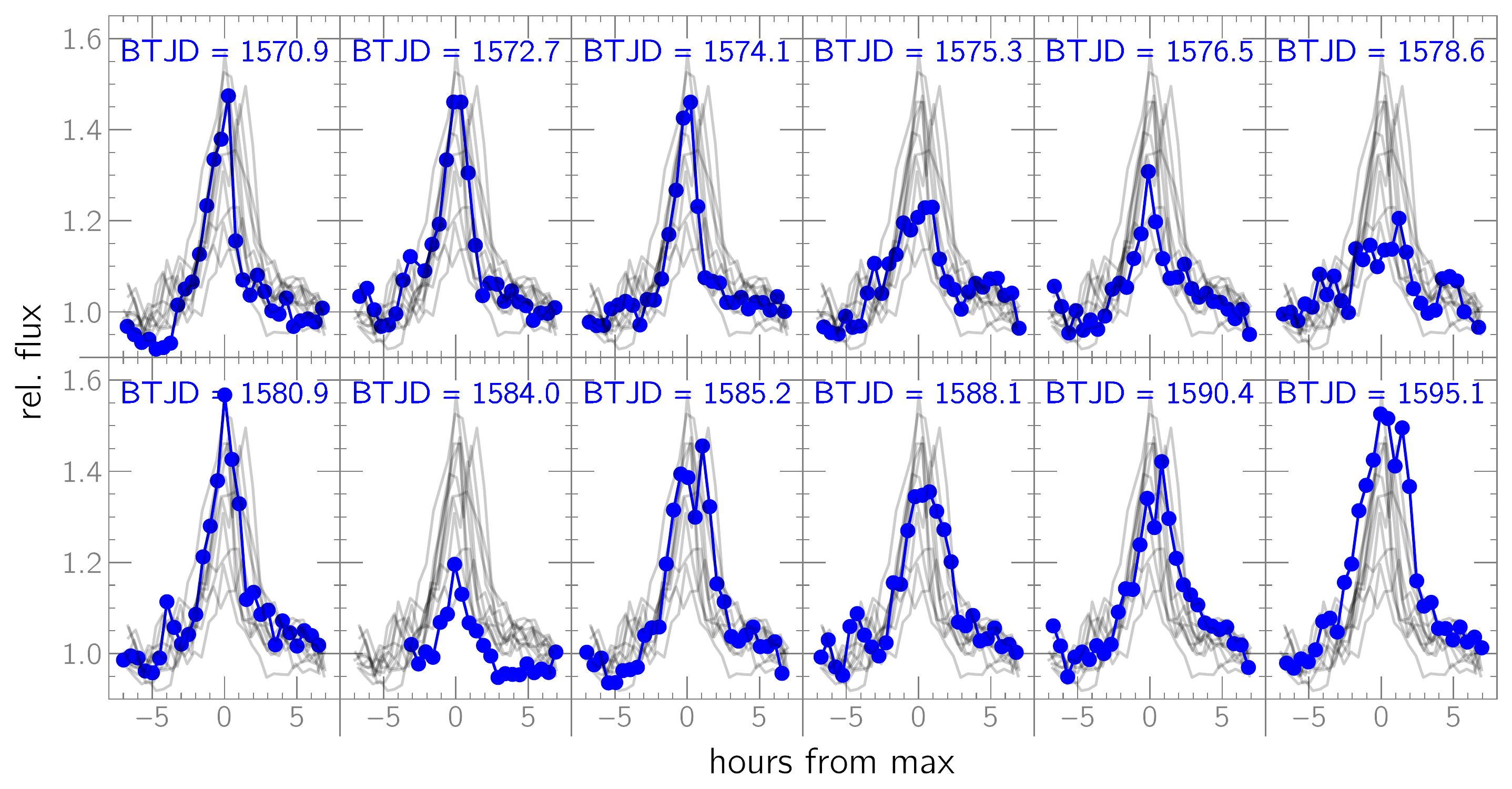}
    \caption{ An enlargement of each outburst (blue markers) from Fig.~\ref{fig:bursts}, superimposed upon the other eleven outbursts (gray lines). The approximate midpoint of each outburst is shown in each panel. Each burst lasted for fewer than six hours, with a mean full-width-at-half-maximum of $3.0\pm1.0$~h.}
    \label{fig:bursts}
\end{figure*}

\section{Data \& Analysis}
    
    \tess\ observed V1025~Cen at the standard 30-minute full-frame cadence between 2019 March 27 and 2019 April 22. Survey photometry from the Catalina Real-Time Sky Survey \citep[CRTS;][]{crts}, the All-Sky Automated Survey for Supernovae \citep[ASAS-SN;][]{shappee, kochanek}, and the Asteroid Terrestrial-impact Last Alert System \citep[ATLAS;][]{tonry, smith} in Fig.~\ref{fig:asassn} establishes that the \tess\ observation occurred during the final recovery from a years-long low state independently identified by Covington et al. (in prep). The ASAS-SN $g$-band and ATLAS $c$-band observations disagree as to the depth of this low state, which is a likely consequence of blending in the ASAS-SN data, which uses a photometric aperture radius of 16~arcsec. The ATLAS All-Sky Stellar Reference Catalog \citep{tonry_catalog} estimates that within a radius of 15.4~arcsec of V1025~Cen, the contaminating $G$-band flux from other sources is equivalent to the $G$-band flux of V1025~Cen at $G=17.1$; if this modeling is reasonably accurate for blending in the $g$~band, it could easily account for the comparatively shallow depth in the ASAS-SN $g$ data.

    We used {\tt TESSCut} \citep{tesscut} and {\tt lightkurve} \citep{lightkurve} to retrieve the full-frame \tess\ observations of V1025~Cen and to extract its background-subtracted light curve with a custom photometric aperture. As a safeguard, we also checked light curves of nearby field stars and the background to ensure that no obvious systematic problems afflicted the data. The most remarkable feature of the resulting light curve (Fig.~\ref{fig:TESS_LC}) is a series of twelve brief bursts, each lasting no longer than $\sim$6~h. The burst durations are all comparable, and by fitting a Gaussian to each burst, we find that their average full-width-at-half-maximum was 3.0$\pm$1.0~h (Fig.~\ref{fig:bursts}). The full duration of each burst was $\lesssim$6~h, but the time between consecutive bursts was irregular, ranging from 1.1~d to 4.6~d. 
    
    Fig.~\ref{fig:TESS_LC} also includes the simultaneous ASAS-SN observations. Due to sampling limitations, only one of the bursts was detected by ASAS-SN. Curiously, the agreement between the ASAS-SN and \tess\ light curves degrades near the end of the observation. In some instances, the ASAS-SN observations show evidence of rapid variation on minutes-long timescales; for example, in a trio of exposures obtained within a 5~min span near BTJD=1594,\footnote{BTJD is defined as BJD$-$2457000.} V1025~Cen's initial flux density was $3.25\pm0.25$~Jy, but the next two exposures showed it to be $1.99\pm0.23$~Jy and $2.40\pm0.31$~Jy. A separate group of observations, obtained hours earlier, also showed variability in excess of the uncertainties. The rapid variations in the ASAS-SN data do not correlate with the full-width-at-half-maximum of the stellar point-spread functions, which suggests that they are not attributable to time-varying contamination from other sources. Furthermore, when the ASAS-SN data show this increased internal variability, the \tess\ light curve becomes visibly jagged. We speculate, therefore, that during portions of the \tess\ observation, V1025~Cen showed rapid variability that could not be resolved at the 30~min cadence of the full-frame data.

    The bursts in the \tess\ light curve have no precedent in the observational literature for V1025~Cen. \citet{buckley98} is the only published study that includes photometry of V1025~Cen, and it reported a snapshot observation at $V \sim 16.1$, along with time-series light curves showing variability between $B\sim15.5-16$ for several nights in 1995 April. There is no evidence of bursts in any of their data.

\begin{figure}
        \centering
        \includegraphics[width=\columnwidth]{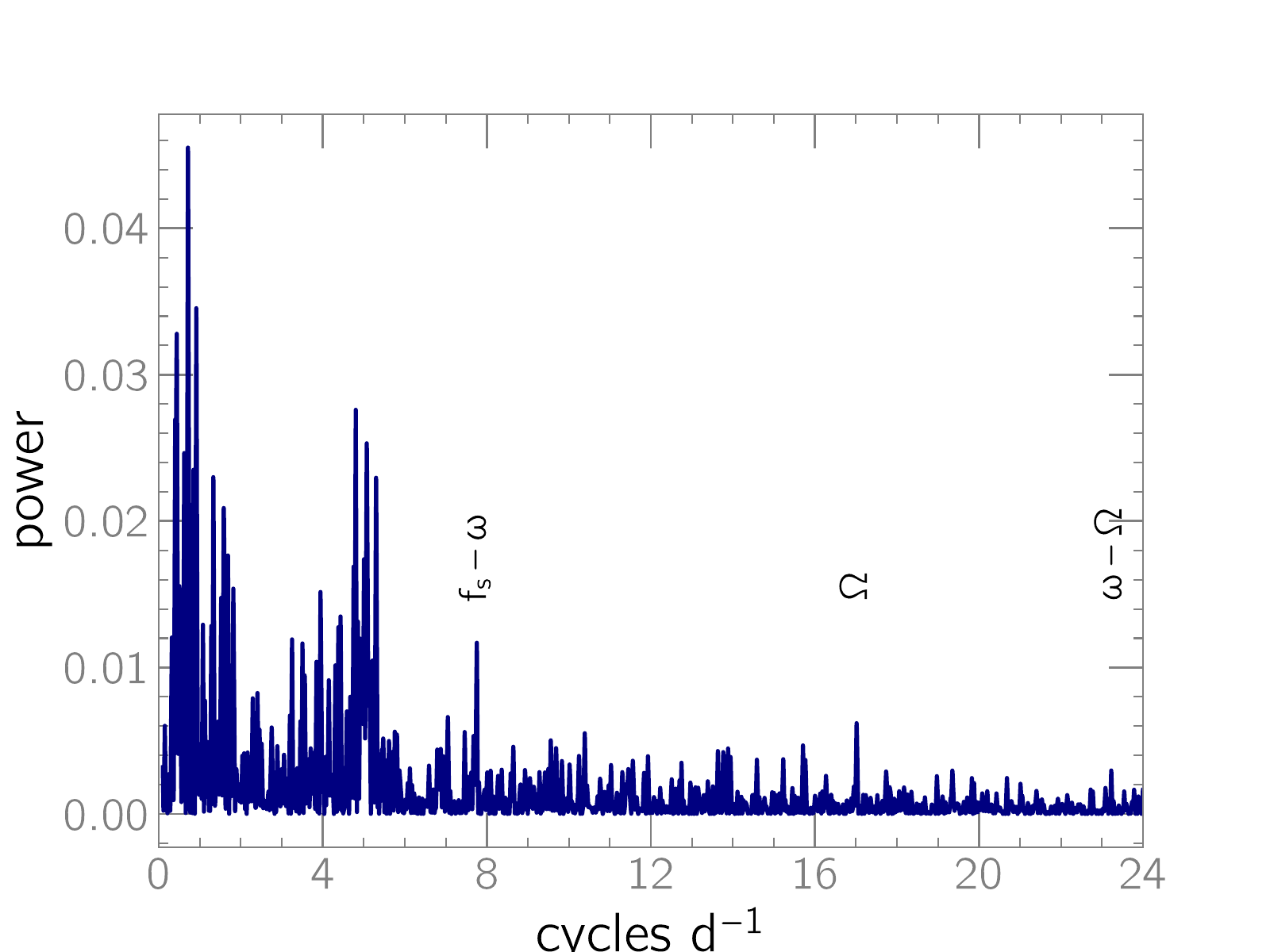}
        \caption{Power spectrum of V1025~Cen during the intervals between the bursts. $f_{s}$ refers to the sampling frequency of 48~cycles~d$^{-1}$, while $\omega$ is the WD spin frequency and $\Omega$ is the binary orbital frequency. Because $\omega$ (40.25~cycles~d$^{-1}$) is above the Nyquist frequency of the data (24~cycles~d$^{-1}$), it produces an alias at a frequency of $f_s - \omega = 7.75$~cycles~d$^{-1}$. The power near 5~cycles~d$^{-1}$ is of unknown origin.}
        \label{fig:power}
    \end{figure}

    Unfortunately, the WD's spin frequency ($\omega$ = 40.25~cycles~d$^{-1}$) is above the Nyquist frequency of the \tess\ data (24~cycles~d$^{-1}$), and the resulting aliasing precludes us from using the power spectrum to ascertain whether the WD was accreting from an accretion disk or directly from the ballistic accretion stream \citep{ferrario99}. Fig.~\ref{fig:power} shows the power spectrum of the light curve during the intervals between the bursts, and while both $\Omega$ and $\omega-\Omega$ are present, there is also a Nyquist-aliased signal at 7.75~cycles~d$^{-1}$, equivalent to $f_{s} - \omega$, where $f_{s}$ is the sampling frequency of 48~cycles~d$^{-1}$. Even more striking is a cluster of signals near 5~cycles~d$^{-1}$, corresponding to a quasi-periodic signal that can be seen directly in the light curve in Fig.~\ref{fig:TESS_LC}, particularly after BTJD$\sim$1590. We were unable to attribute these signals to Nyquist aliasing of $\omega$, $2(\omega-\Omega)$, or other super-Nyquist frequencies; moreover, Nyquist aliasing of a periodic signal should not convert it into a cluster of aliases. It is conceivable, however, that a quasi-periodic oscillation above the Nyquist frequency could generate a cluster of low-frequency signals with no discernible relation to V1025~Cen's periodic variations.

\section{Discussion}

    \subsection{Magnetic gating in V1025 Cen}
    
    \citet{hl17} found that for realistic disk viscosities, the DN instability cannot produce the hours-long bursts observed in some IPs. They instead suggested that these bursts might be due to some instability coupling the white dwarf magnetic field with that generated by the magnetorotational instability operating in the accretion disk. In particular, they mention the \citet{st93} instability, which was suggested by \cite{Mhlahlo07} to explain EX~Hya's outbursts, but as \citet{Mhlahlo07} pointed out, this explanation struggled to account for the long recurrence times between that system's outbursts.
    
    This difficulty is not present with the highly repetitive bursts of V1025~Cen, which we can readily explain using the \citet{st93} instability. \citet{DS10,DS11,DS12} applied a model based on the \citet{st93} mechanism to outbursts in millisecond pulsars and young stellar objects (YSOs), and the light curve of V1025~Cen shows strong similarities to the synthetic lightcurves of \citet{DS10,DS11,DS12} as well as the Type~II bursts of the accreting neutron star known as the Rapid Burster. These similarities make the \citet{st93} instability an excellent candidate to be the cause of the bursts in V1025~Cen, a hypothesis that we examine quantitatively below.  
    
    The applicability of the magnetic gating model to V1025~Cen is strongly supported by the parameters of this system. Two important radii are the corotation radius, defined as \begin{equation}
    R_{\mathrm{cor}}=3.52 \times 10^{10} M_{1}^{1 / 3} P_{\rm spin(h)}^{2/3} \mathrm{~cm},
    \end{equation}
    and the magnetospheric radius, given by
    \begin{equation}
    R_{\mathrm{M}}=1.4 \times 10^{10} \dot{M}_{15}^{-2 / 7} M_{1}^{-1 / 7} \mu_{32}^{4 / 7} \mathrm{~cm}.
    \end{equation} In these equations, $M_1$ is the mass of the WD in solar masses, $\dot{M}$ is the mass-transfer rate in units of $10^{15}$~g~s$^{-1}$, and $\mu$ is the magnetic moment in units of $10^{32}$~G~cm$^3$.
    For V1025~Cen, $R_{\mathrm{cor}}=2.5 \times 10^{10}M_{1}^{1 / 3} $~cm, and to have $R_{\mathrm{cor}}\approx R_{\mathrm{M}}$ the magnetic moment should be
    \begin{equation}
    \mu \approx 2.9 \times 10^{32} \dot{M}_{15}^{1/2} M_{1}^{5/ 6} P_{\rm spin(36)}^{7/6} \mathrm{G\,cm^3},
    \end{equation}
    where $P_{\rm spin(36)}$ is the spin period expressed as a fraction of the observed 36-min spin period of V1025~Cen. V1025~Cen satisfies the necessary condition for the instability ($R_{\mathrm{cor}}\approx R_{\mathrm{M}}$)
    if its accretion rate is close to the expected secular value \citep[$\dot{M}_c \sim 10^{15}$g~s$^{-1}$;][]{kbp} and if the WD's magnetic moment is typical of the assumed magnetic moments in IPs.
    
    The fact that outbursts appear when the system is at its brightest in Fig.~\ref{fig:asassn} suggests that in the preceding epochs, the accretion rate was less than the critical value above which the disk would become thermally unstable and subject to the DNI. Alternatively, the lack of outbursts during the low state might suggest that the accretion disk dissipated \citep{hl17_lowstates}.

    Although \citet{st93} and \citet{DS10, DS11, DS12} specifically considered a truncated accretion disk, their modeling is not predicated upon a Keplerian accretion flow. Instead, what is crucial for the mechanism to work is a ring of matter accumulated near the magnetospheric/corotation radius. A Keplerian disk would fulfill this requirement, but even if V1025~Cen were diskless during the \tess\ observation, modeling by \citet{kw99} and \citet{norton} shows that diskless accretion flows can take the form of a torus that surrounds the magnetosphere.
    
    It is unclear whether a disk exists in V1025~Cen. The circularization radius of the binary is \citep[using][]{GA05}
    \begin{equation}
    R_{\mathrm{circ}}=2.6 \times 10^{9} M_{1}^{1 / 3}q^{-0.48} \left( 1 + q \right)^{0.57}P_{\rm orb(h)}^{2/3} \mathrm{~cm},
    \end{equation}
    which for V1025 Cen is $R_{\mathrm{circ}}= 1.0\times 10^{10}$cm, assuming $q=0.1$.
    Therefore, $R_{\mathrm{circ}} \lesssim R_{\mathrm{cor}}$, and a Keplerian disk is unlikely unless $q$ is very small. Moreover, observations by \citet{hellier02} were inconclusive on the existence of a disk, and in any event, they would not resolve the issue of whether one was present nearly two decades later during the \tess\ observation. On one hand, the unambiguous existence of a Keplerian disk would simplify the application of the \citet{st93} mechanism to V1025~Cen because it would establish the presence of the ring of matter required by that instability. On the other hand, the arguments presented by \citet{st93} and \citet{DS10,DS11,DS12} appear to be generalizable to the torus-like accretion flows expected to form when  $R_{\mathrm{cor}} \sim R_{\mathrm{M}}$, a regime in which the kick provided by magnetic forces is insufficient to expel matter to large distances. However, it will be important for theoretical studies to test our proposed generalization of the \citet{st93} instability.

    As for the outburst timescales, they depend strongly not only on $\dot M_c$, but also on the two transition widths $\Delta R/R_{\rm M}$ (for the torque) and $\Delta R_2/R_{\rm M}$ \citep[for the mass flux;][]{DS11}, so it is not possible to use the simulations for neutron stars or young stellar objects (YSOs) to scale them to the case of IPs. For example, the recurrence time in the simulations vary from 1 to 1000 times the viscous time at the magnetospheric radius.

     Figures~\ref{fig:asassn}~and~\ref{fig:TESS_LC} establish that the WD continues to accrete between bursts; if the accretion rate were negligible, V1025~Cen would have faded to at least the brightness level observed during its low state. \citet{DS12}, in a regime that they referred to as ``RII,'' found that the \citet{st93} mechanism can allow for reduced---but uninterrupted---accretion when the transition region between accreting and non-accreting states is large. Bursts in this regime have comparatively smooth profiles and short recurrence timescales compared to those in the \citet{DS12} RI regime, whose bursts have distinctive initial spikes and long intervals of negligible accretion. Moreover, the RII instability from \citet{DS12} occurs only for a limited range of $\dot{M}$, which agrees well with the apparent sensitivity of the bursts to the accretion rate (see Sec.~\ref{disc:asassn}). However, the generalization of the instability to a torus-like geometry is an important step that still requires theoretical confirmation.

    In the model, the outburst amplitudes are independent of the parameters and roughly a factor two of $\dot M_c$, which seems to be compatible with observations. On the other hand, it is unclear whether this independence of parameters is a generic property of the model.

    \subsection{Magnetic gating in other IPs}
    
    Two studies have applied the \citet{st93} model to {\it suspected} magnetic systems. \citet{scaringi} found that MV~Lyr shows recurring bursts of accretion every  $\sim2$~h, each lasting for $\sim30$~min, at its lowest accretion rates. MV~Lyr would have to possess a substantially lower magnetic-field strength \citep[$\sim0.02-0.13$~MG;][]{scaringi} than what is normally inferred for IPs \citep[$\sim1-10$~MG;][]{patterson}, including V1025~Cen. The short duration ($\sim30$~min) and recurrence time ($\sim2$~h) in MV~Lyr compared to V1025~Cen would be accounted for by the weakness of the magnetic field, leading to a small $R_{\rm M}$ and hence a small viscous time at the inner edge of the disk. 
   
   Likewise, with TW~Pic, magnetically gated bursts occur at diminished accretion rates, typically last for $\sim30$~min, and recur every $1.2-2.4$~h \citep{scaringi21}. As with MV~Lyr, the short durations and recurrence intervals of TW~Pic's bursts are consistent with it possessing a lower field strength than V1025~Cen. \citet{scaringi21} constrained the field strength to be $\lesssim1$~MG but found that it is degenerate with the unknown spin period of the WD. 
    
  Because V1025~Cen has already been independently confirmed to be an IP \citep{buckley98, hellier98}, the detection of magnetic gating in its \tess\ light curve provides important validation of the magnetic-gating interpretation by \citet{scaringi, scaringi21} and its extension to magnetized WDs in general. Additionally, because accreting WDs are more common than accreting neutron stars, the presence of magnetic gating in at least some of these systems offers a convenient means of amassing observations against which theoretical predictions can be tested.

\subsection{Difficulty of detecting the bursts in survey photometry}

    \label{disc:asassn}
    
    Ground-based photometric surveys such as ASAS-SN have proven adept at detecting the presence of days- or weeks-long outbursts in non-magnetic systems. In contrast, the comparison of the \tess\ and ASAS-SN light curves in Fig.~\ref{fig:TESS_LC} shows how easily V1025~Cen's rapid bursts can elude the relatively frequent ASAS-SN observations. Even though there were a dozen bursts in just four weeks in the \tess\ light curve, they are inconspicuous in the ASAS-SN photometry because of the combination of the survey's sampling and the bursts' low amplitudes and short durations. Only one of the twelve bursts in the \tess\ light curve was also observed by ASAS-SN, and that particular burst is an inconspicuous feature in the long-term ASAS-SN light curve in Fig.~\ref{fig:asassn}. Were it not for the simultaneous \tess\ data, it could easily be dismissed as strong flickering or even a data-quality artifact.
    
    Such a low detection efficiency has several important implications. For example, the ASAS-SN light curve shows at least four additional short-lived, low-amplitude brightenings in the months surrounding the \tess\ observation during the 2018-2019 observing season for V1025~Cen. These features resemble the lone \tess\ burst detected by ASAS-SN. If we assume that these are the same phenomenon observed by \tess, the presence of several bursts in the ASAS-SN light curve implies that many more eluded detection during that timespan. Therefore, we infer that the bursts in the \tess\ light curve were a common feature in the months surrounding the \tess\ observation, when V1025~Cen was finishing its recovery from a years-long low state. 
    
    A simple statistical analysis of the ASAS-SN data bolsters this inference. When it observes a target, ASAS-SN usually obtains multiple photometric measurements over the course of a few minutes, so we created a new light curve in which each unique visit was represented by a single point whose abscissa was the median time of observation and whose ordinate was the median magnitude during that visit. Using this binned light curve, we analyzed each observing season separately by detrending that season's light curve and counting the number of measurements that were $>3\sigma$ brighter than the mean magnitude. The 2018-2019 observing season contained four such measurements; for comparison, a Gaussian distribution with the observed N=241 would be expected to show just 0.3 measurements more than $3\sigma$ below the mean. Moreover, of the nine observing seasons covered by ASAS-SN, only two others showed even a single bright, $>3\sigma$ outlier. It therefore appears likely that the bursts occurred preferentially at the end of the low state in 2018 and 2019. This, in turn, provides observational evidence that magnetic gating in V1025~Cen is very sensitive to the accretion rate.

    \section{Conclusion}
    
    The \textit{TESS} light curve of V1025~Cen contains a dozen bursts of accretion that are notable for their extremely short durations ($\lesssim$6~h) and rapid recurrence timescale ($\sim1-3$~d). We show that the instability proposed by \citet{st93} and expanded in \citet{DS10, DS11, DS12} can explain these bursts as episodes of magnetically gated accretion. Although \citet{scaringi, scaringi21} have identified magnetic gating during low-accretion states in MV~Lyr and TW~Pic, respectively, neither of these systems is a confirmed IP. Conversely, the WD in V1025~Cen was already known to be magnetic, which bolsters the application of the magnetic gating model to accreting WDs and beyond its original context of accreting neutron stars and YSOs.

    The duty cycle and amplitudes of the bursts in V1025~Cen are sufficiently low that they are difficult to recognize in survey photometry, despite reasonably good coverage. It is possible that other IPs show similar bursts, and if they do, sustained time-series photometry will be necessary to detect them unambiguously, as the \tess\ light curve of V1025~Cen vividly demonstrates.
    
\acknowledgments

We thank the anonymous referee for a helpful report that improved this manuscript.

PS and CL acknowledge support from NSF grant AST-1514737.

JPL's research was supported by a grant from the French Space Agency CNES.

M.R.K. acknowledges funding from the Irish Research Council in the form of a Government of Ireland Postdoctoral Fellowship (GOIPD/2021/670: Invisible Monsters).

\software{{\tt astropy} \citep{astropy}, {\tt lightkurve} \citep{lightkurve}, {\tt TESScut} \citep{tesscut}}

\bibliography{bib.bib}

\end{document}